\documentclass[aps,prb,amsmath,twocolumn,amssymb,notitlepage,superscriptaddress]{revtex4-1}	
\usepackage[english]{babel}
\usepackage{graphicx}
\usepackage{amsmath}
\usepackage{amssymb}
\usepackage{epstopdf}
\usepackage{amsfonts}
\usepackage{bm}
\usepackage{mathdots}
\usepackage{times}
\usepackage{mathtools}
\usepackage{hyperref}

\usepackage{color}

\newcommand{\astcycl}{\mathrlap{\kern0.085em{\circlearrowright}}\ast}
\newcommand{\taucycl}{\mathrlap{\kern0.42em{\bullet}}\circlearrowright}

\newcommand{\notilde}{}

\newcommand{\CFMethod}{$\mathcal{O}_0$}
\newcommand{\CMethod}{$\mathcal{O}_2$}

\begin{document}
\title{Simulation of time-dependent resonant inelastic X-ray scattering \\ using non-equilibrium dynamical mean-field theory}
\author{Martin Eckstein}
\affiliation{Department of Physics, University of Erlangen-N\"urnberg, 91058 Erlangen, Germany}
\author{Philipp Werner}
\affiliation{Department of Physics, University of Fribourg, 1700 Fribourg, Switzerland}

\begin{abstract}
We develop a framework to evaluate the time-dependent resonant inelastic X-ray scattering (RIXS) signal with the use of non-equilibrium dynamical mean field theory simulations. The approach is based on the solution of a time-dependent impurity model which explicitly incorporates the probe pulse. It avoids the need to compute four-point correlation functions, and can in principle be combined with different impurity solvers. This opens a path to study time-resolved RIXS processes in multi-orbital systems. The approach is exemplified with a study of the RIXS signal of a melting Mott antiferromagnet. 
\end{abstract}

\maketitle

\section{Introduction}

Over the past years, time-resolved spectroscopies have evolved into a powerful framework for probing the ultrafast light-induced dynamics in complex solids.\cite{Basov2017, Claudio2016} A  potentially very versatile variant is resonant inelastic X-ray scattering (RIXS).\cite{Ament2011}  In RIXS, an incoming photon with energy $\omega_{\text{in}}$ and momentum $\bm q_{\text{in}}$ is scattered into an outgoing photon with energy $\omega_{\text{out}}$ and momentum $\bm q_{\text{out}}$ via an intermediate state which involves a short-lived core hole. The energy loss $\omega_{\text{in}}-\omega_{\text{out}}$ measures the excitation spectrum in the solid. Because the core level is strongly localized, the excitation is element-specific, and the polarization of the photons can be used to achieve orbital selectivity. The measurement is therefore ideally suited to probe the charge, spin, and orbital degrees of freedom in complex solids.\cite{Hill1998, Braicovich2009, Schlappa2012,Johnston2016,Schlappa2018} Free-electron lasers can provide the necessary brilliant and ultrashort X-ray pulses to acquire femtosecond time-resolution in pump-probe experiments, where a short X-ray  pulse acts on the sample around a given probe time $t_p$. This makes time-resolved RIXS a very promising technique for revealing the ultrafast dynamics of entangled degrees of freedom in solids. \cite{Dean2016,Mitrano2019,Mitrano2020}

The interpretation of the RIXS signal relies on a microscopic understanding of the intermediate state dynamics of the solid in the presence of a core hole. The scattering  amplitude from an initial state $|\Psi_0\rangle$ at time $t_0$ into a final state $|\Psi_f\rangle$ at time $t$ with a photon in the outgoing mode is given by all processes which involve the absorption and emission of a photon at consecutive times $t_1$ and $t_2$,
\begin{align}
\langle \Psi_f|\mathcal{U}(t,t_2) P_{\text{out}}(t_2)\mathcal{U}(t_2,t_1)P^\dagger_{\text{in}}(t_1)\mathcal{U}(t_1,t_0)|\Psi_0\rangle.
\label{4point}
\end{align}
Here $P_{\nu}^\dagger$ ($P_{\nu}$) is the core-valence dipolar  transition operator for the ingoing ($\nu=\text{in}$) and outgoing ($\nu=\text{out}$) photon mode, and $\mathcal{U}$ the time-evolution operator. The RIXS signal is related to the square of this amplitude, and is therefore linked to a four-point correlation function in time.\cite{Kramers1925} The evaluation of the latter is 
challenging in an extended solid, in particular within the framework of many-body perturbation theory, where it requires  the infinite re-summation of diagrams into so-called vertex corrections. RIXS spectra are therefore typically computed using exact diagonalization on 
small clusters. While such cluster schemes are very successful in many respects,\cite{Ament2011, Norman2018} they cannot easily incorporate the itinerant nature of the conduction electrons in solids. This difficulty may be overcome within the framework of dynamical mean field theory (DMFT),\cite{Georges1996} as proposed by Hariki and co-workers.\cite{Hariki2018, Hariki2020}  Within DMFT, local properties of an atom in the solid are obtained from an impurity model, where the atom is coupled to a self-consistently determined particle reservoir. It is therefore natural to evaluate the RIXS signal in this impurity model (similar as done for X-ray absorption spectroscopy\cite{Haverkort2014}), which incorporates the feedback of the itinerant degrees of freedom on the atom.  This approach does not yet resolve the momentum transfer to the photon, but it captures the important dependence of the signal on the energy loss and light polarization (orbital selectivity) even in the single-site DMFT framework. Momentum resolution could be gained within cluster extensions of DMFT.\cite{Maier2005}

The description of time-resolved measurements on systems out of equilibrium is even more involved. A real-time formulation of the classic Kramers expression \cite{Chen2019}  has been investigated successfully for a Hubbard model using  matrix-product states in one-dimension,\cite{Zawadzki2020} and using exact diagonalization for clusters of a two-dimensional model.\cite{Wang2020}  Here, we discuss a generalization of the DMFT evaluation of RIXS to non-equilibrium situations. Again, this approach has the advantage that it incorporates the itinerant degrees of freedom. More importantly,  the  
impurity model in non-equilibrium DMFT  self-consistently incorporates changes of the local electronic structure of the probe site which result from the non-equilibrium excitation of the whole solid.\cite{Aoki2014} Finally,  diagrammatic techniques to solve the quantum impurity model out of equilibrium, such as a the strong coupling hybridization expansion,\cite{Eckstein2010nca} allow to treat multi-orbital impurity models, which would require an exponentially large Hilbert space in exact diagonalization. 

A straightforward diagrammatic evaluation of RIXS spectra within non-equilibrium (cluster) DMFT would rely on 
the explicit evaluation of four-point time correlation functions of the impurity model, including vertex corrections which are in essence processes in which the impurity emits an electron into the reservoir {\em before} the core hole creation (annihilation) and absorbs another one {\em after} that. The systematic re-summation of such vertex corrections for out-of equilibrium Green's functions is beyond reach of the present numerical techniques. To circumvent this problem, we here implement an explicit time-dependent impurity model which reproduces the RIXS spectrum. We investigate two variants, one which relies on the evaluation of a two-point correlation function, and one which explicitly includes all relevant states of the photon mode and thus allows to directly measure the signal from the photon occupation. A comparison of the two approaches allows to judge the effect of vertex corrections in the two-point correlation function, and therefore the reliability of the results. The formalism is benchmarked on several problems, including the dynamics in an antiferromagnetic Mott insulator out of equilibrium. Further applications to multi-orbital models will be presented elsewhere.\cite{Werner_RIXS}

The paper is structured as follows. In Sec.~\ref{sec:model} we introduce the DMFT impurity model from which the RIXS spectrum is obtained, and in Sec.~\ref{sec:rixs} we explain the formalism. Section~\ref{sec:bench} contains benchmarks and applications of the formalism for an impurity model with a single bath orbital (Sec.~\ref{sec:bench:one}), a single impurity Anderson model (Sec.~\ref{sec:bench:siam}), and the DMFT solution for the dynamics in an antiferromagnetic Mott insulator (Sec.~\ref{sec:bench:afm}). Finally, Sec.~\ref{sec:conclusion} contains a conclusion and outlook.

\section{Model}
\label{sec:model}

\subsection{Local Hamiltonian}

To evaluate the RIXS amplitude, we start from a generic Hamiltonian for one atomic unit in the solid,
\begin{align}
\label{hloc}
H_\text{loc}=
H_d 
+
H_c
+
H_{cd}
+
H_\text{ph}
+
H_\text{dip}.
\end{align}
Here $H_d$ describes the valence orbitals, with fermion operators $d_{\gamma,\sigma}$ for orbital $\gamma$ and spin $\sigma$. The local interaction between the electrons in these valence orbitals is arbitrary within our formalism and need not be specified at the moment. Analogously, $H_c$ is the Hamiltonian of the core level(s). While all expressions below can be extended straightforwardly to a general multi orbital core manifold on a formal level, 
we restrict the discussion to a single orbital with energy $\epsilon_c$, and creation (annihilation) operators $c_{\sigma}^\dagger$ ($c_\sigma$),
\begin{align}
H_{c}
=\epsilon_{c}\sum_{\sigma} c_{\sigma}^\dagger c_{\sigma}.
\end{align}
In addition, we describe the core-valence interaction $H_{cd}$ as a density-density interaction
\begin{align}
H_{cd} = U_{cd} \sum_{\sigma,\sigma',\gamma} (c_{\sigma}^\dagger c_{\sigma}-1)d_{\gamma,\sigma'}^\dagger d_{\gamma,\sigma'},
\end{align}
which vanishes in the absence of a core hole. To the electronic model we add the photon modes. The electromagnetic field is restricted to two  modes labelled by $\nu$, i.e., the ingoing ($\nu=\text{in}$)  and outgoing ($\nu=\text{out}$) photon, which are characterized by their energy $\omega_{\nu}$, their polarization $\bm \epsilon_{\nu}$, and their wave vector $\bm q_{\nu}$. Its free Hamiltonian is
\begin{align}
H_\text{ph}
=
\sum_{\nu=\text{in},\text{out}}
\omega_{\nu} b_{\nu}^\dagger b_{\nu}.
\end{align}
 The transverse electric field of each mode at the position $\bm R$ of the atom is  $i\bm \epsilon_\nu (b_{\nu}\eta_{\nu}-b_{\nu}^\dagger\eta_{\nu}^*)$, with $\eta_{\nu}\propto e^{i\bm q_{\nu}\bm R}$, and the light-matter interaction is due to a dipolar coupling
\begin{align}
\label{hidip}
H_\text{dip}
=
\sum_{\nu}
i(b_{\nu}\eta_{\nu}-b_{\nu}^\dagger\eta_{\nu}^*)  
\sum_{\gamma,\sigma} 
(p_{\gamma}^{\nu}  
P_{\gamma,\sigma}^\dagger + h.c.).
\end{align}
Here $P_{\gamma,\sigma}^\dagger=d_{\gamma,\sigma}^\dagger c_{\sigma}$ creates a core hole by transferring the electron to orbital $\gamma$, and $p_{\gamma}^{\nu}=\langle d_{\gamma}| \bm\epsilon_{\nu}\cdot  \bm r  | c \rangle$ is the dipolar transition matrix element for the mode with polarization $\bm\epsilon_{\nu}$. This concludes the definition of the local Hamiltonian. In the following we use the abbreviation
\begin{align}
P_{\nu,\sigma}^\dagger=
\sum_{\gamma} 
p_{\gamma}^{\nu}   P_{\gamma,\sigma}^\dagger.
\end{align}
Moreover, for the evaluation of the RIXS spectrum the incoming photon in Eq.~\eqref{hidip} is replaced by a classical driving field
\begin{align}
\label{field}
b_{\text{in}}
\to 
s(t)e^{-i\omega_{\text{in}} t },
\end{align}
where $s(t)$ the the probe envelope that defines the time window in which the probe is acting on the solid. The operators $b_{\text{in}}$ and $b_{\text{in}}^\dagger$  therefore do no longer appear explicitly, and we 
will omit the index $\text{out}$ for the outgoing photon operators in the following.  

\subsection{Rotating wave approximation}

Usually, there is a large energy  scale separation between the relevant energy transfer $\omega_{\text{in}}-\omega_{\text{out}}$ to the valence band, which is at most of the order of few eV, and the absolute energies $\omega_{\text{in}}$, $\omega_{\text{out}}$, and $|\epsilon_{c}|$, which can be of the order of  $1000$ eV. One can therefore assume that $\omega_{\nu}=\tilde \omega_{\nu}+E$,  $\epsilon_c=-E+\tilde \epsilon_c$, with $\tilde \omega_{\nu},\tilde\epsilon_c\ll E$.  In this limit, only the terms $b_{\nu} P_{\nu,\sigma}^\dagger$ and $b_{\nu}^\dagger P_{\nu,\sigma}$, which either create a core-hole under the absorption of a photon, or annihilate a core-hole under photon emission, should be of relevance. This is made mathematically rigorous by the rotating wave approximation, which is recapitulated in the following.
 
The Hamiltonian $H_\text{loc}$ is first  rewritten using a canonical transformation which shifts the energy of the core band and the photon. In general, after a time-dependent basis transformation 
$|\tilde \psi\rangle = \mathcal{W}(t) |\psi\rangle$ the new wavefunction satisfies the Schr\"odinger equation
$i\partial_t |\tilde \psi\rangle = \tilde H| \tilde \psi\rangle$ with $\tilde H=\mathcal{W}H\mathcal{W}^\dagger +i(\partial_t \mathcal{W})\mathcal{W}^\dagger$. The choice $\mathcal{W}=\exp\big[itE\big(b^\dagger b-\sum_{\sigma}c_{\sigma}^\dagger c_{\sigma} \big) \big]$ transforms the operators in the  dipolar coupling Eq.~\eqref{hidip}  like
$b^\dagger P^\dagger \to b^\dagger P^\dagger e^{-i2Et}$, $b P \to b P e^{i2Et}$, $b P^\dagger \to b P^\dagger$, $b^\dagger P \to b^\dagger P$, and shifts the core and photon energies to $ \tilde\epsilon_{c}$ and $\tilde \omega_{\text{out}}$. The rotating wave approximation then amounts to taking $E\to\infty$, so that all terms oscillating with $e^{\pm i2Et}$ vanish. The resulting Hamiltonian is 
\begin{align}
\label{hloc_RWA}
\tilde H_\text{loc} 
&=
\tilde H_d 
+
\tilde H_c
+
\tilde H_{cd}
+
\tilde H_\text{ph}
+
\tilde H_{\text{in}}
+
\tilde H_{\text{out}},
\end{align}
where the valence Hamiltonian ($\tilde H_{d}=H_{d}$) and the core-valence interaction are unchanged ($\tilde H_{cd}=H_{cd}$), and
\begin{align}
&\tilde H_\text{ph}+\tilde H_{c}
=\tilde \omega_{\text{out}} b^\dagger b+\tilde \epsilon_{c}\sum_{\sigma} c_{\sigma}^\dagger c_{\sigma} 
\\
\label{ingoung}
&\tilde H_{\text{in}}
=
\sum_{\sigma}
\big[i\eta_{\text{in}}s(t) e^{-i\tilde \omega_{\text{in}} t }P_{\text{in},\sigma}^\dagger  +h.c.\big],
\\
\label{outgoung}
&\tilde H_{\text{out}}
=
\sum_{\sigma}  
\big[i\eta_{\text{out}}b P_{\text{out},\sigma}^\dagger
+h.c.\big].
\end{align}
In the following, we work with the Hamiltonian in the rotating wave approximation, and from now on omit the tilde for $\tilde \omega_{\nu}$; $\tilde \epsilon_c$ can be set to zero without loss of generality. 

\subsection{DMFT embedding}

The atomic model \eqref{hloc_RWA}  is now embedded in a lattice. Within DMFT, the lattice is replaced by an impurity model,\cite{Georges1996} where the local Hamiltonian \eqref{hloc_RWA} is coupled to a self-consistently determined particle reservoir. Because we aim to study real-time processes, we work within the Keldysh formalism with a closed time contour $\mathcal{C}$.\cite{Aoki2014} The  impurity model is represented through its action
\begin{align}
\label{simp}
\mathcal{S}_\text{imp} =
\mathcal{S}_\text{loc}
+
\mathcal{S}_{dd}
+
\mathcal{S}_{cc},
\end{align}
where the local action is simply defined by Hamiltonian \eqref{hloc_RWA},
\begin{align}
\mathcal{S}_\text{loc} = -i \int_{\mathcal{C}} dt  H_\text{loc}(t)
\end{align}
and the second term with the so-called hybridization function $\Delta_{\gamma,\gamma'}(t,t')$,
\begin{align}
\label{sdd}
\mathcal{S}_{dd} = -i \int_{\mathcal{C}} dt dt' \,\sum_{\sigma,\gamma,\gamma'} d_{\gamma,\sigma}^\dagger (t) \Delta_{\gamma,\gamma'}(t,t') d_{\gamma',\sigma}(t'),
\end{align}
describes the hybridization of the valence orbitals with the self-consistent bath. Within DMFT, the impurity can exchange electrons with the bath during the RIXS process and the intermediate state evolution, but because the core-valence interaction is local, the electronic structure of the bath is not affected by the RIXS process itself.\cite{Hariki2018} In the time-dependent formalism, the hybridization function $\Delta_{\gamma,\gamma'}(t,t')$ is therefore computed in a separate non-equilibrium DMFT simulation, which includes the pump laser fields or other non-equilibrium excitations that drive the system out of equilibrium without core hole excitation. The RIXS signal is then subsequently evaluated from the impurity model \eqref{simp} with given $\Delta_{\gamma,\gamma'}(t,t')$. 

In addition, in the action \eqref{simp} a particle reservoir is coupled to the core level, 
\begin{align}
\mathcal{S}_{cc} = -i \int_{\mathcal{C}} dt dt' \sum_{\sigma} c_{\sigma}^\dagger (t) \Delta_c(t,t') c_{\sigma}(t'),
\label{bathcc}
\end{align}
which takes the form of a core hybridization function $\Delta_c(t,t')$ analogous to $\Delta_{\gamma,\gamma'}(t,t')$. This core environment should represent an entirely filled reservoir of electrons, so that it can lead to the decay of a core-hole, but not its creation.  Within the Keldysh formalism, a filled bath is realized by a hybridization function $\Delta_{c}$ for which the unoccupied density of states (greater component) vanishes.\cite{Aoki2014}Introducing this bath adds a lifetime to the core hole, and therefore takes a role analogous to the broadening $\Gamma$ of the intermediate states introduced in the exact-diagonalization formalism.\cite{Chen2019} In the simulations, we use an analytic form corresponding to a Gaussian density of states $\Delta_{c}(\omega)$ of bandwidth $W$, so that 
\begin{align}
\Delta_{c}^{\text{ret}}(t-t')
&=-i\theta(t-t')\int d\omega \Delta_{c}(\omega)e^{-i\omega(t-t')}\nonumber
\\
\label{cbath}
&=-i  \theta(t-t')2\pi^{-1/2} W\Gamma e^{ -(t-t')^2W^2}.
\end{align}
The bath is normalized such that an isolated core-hole would be filled within a lifetime $\Gamma^{-1}$. The bandwidth is made  as large as possible, so that details of the bath density of states become unimportant (numerically, one has to resolve the time-dependence of $\Delta_{c}(t-t')$ within the given time grid).

\section{Evaluation of the RIXS signal}
\label{sec:rixs}

\subsection{Overview}

In this section, we evaluate the RIXS signal for the time-dependent impurity model defined by the action \eqref{simp}. Before the pulse $s(t)$ is applied, the system is in a product state of  some valence band state, filled core orbitals, and an empty outgoing photon mode. The RIXS spectrum is then given by the photon occupancy in the long-time limit,
\begin{align}
\label{gegdh}
I_{\text{RIXS}}(\notilde \omega_{\text{out}},\notilde \omega_{\text{in}})
=
\lim_{t\to\infty} \langle b^\dagger b \rangle_t,
\end{align}
computed to leading order in the dipolar matrix elements, i.e., to second order in $p_\gamma^{\nu}$ for both $\nu=\text{in},\text{out}$. In the following we  first write down the standard perturbative expression for $I_{\text{RIXS}}(\notilde \omega_{\text{out}},\notilde \omega_{\text{in}})$, which recapitulates the formulation of the Kramer's formula in real time.\cite{Chen2019} We then discuss how this expression can be evaluated without explicitly computing a four-point correlation function in time, by incorporating time-dependent source fields explicitly in the model. We implement  two possible formalisms, which will be denoted as the \CFMethod{}  and \CMethod{}  approach: 
\begin{itemize}
\item[(1)]
The \CMethod{}   approach, presented in Sec.~\ref{sec:forma:O2}, explicitly includes the ingoing RIXS pulse \eqref{ingoung} in a time-dependent simulation of the model, and will evaluate the RIXS signal in terms of a two-point polarizability of the driven model.
\item[(2)]
The \CFMethod{}  approach, presented in Sec.~\ref{sec:forma:O0}, includes both the ingoing RIXS pulse \eqref{ingoung}   and the relevant states related to the outgoing photon mode in a time-dependent simulation of the model, and computes the expectation value \eqref{gegdh} directly, without evaluating any high-order correlation function. 
\end{itemize}
While both formalisms are equivalent on a formal level, they become inequivalent when vertex corrections are missing in the evaluation of the two-point correlation function within the \CMethod{}  approach (see discussion in Sec.~\ref{sec:forma:nca}). The \CFMethod{}  approach, while numerically more costly, can therefore be used to quantify the 
importance of the missing vertex corrections in the \CMethod{}  approach. 

\subsection{Fourth order perturbation theory}

To implement the perturbation theory in the light-matter interaction, we represent the dipolar matrix elements as $p_{\gamma}^\nu \to g p_{\gamma}^\nu $, where the factor  $g$ is a formal small parameter which facilitates the power counting and will be set to $g=1$ in the final result. We formulate the perturbation theory in a Hamiltonian language, where $\mathcal{H}$ represents a Hamiltonian including the impurity and the bath. As in standard time-dependent perturbation theory, one separates $\mathcal{H}=H_0(t)+gH'(t)$, where $H'$ contains all terms proportional to $g$ (i.e., the driving Hamiltonian and the coupling to the outgoing photon), and expands the time evolution operator as $\mathcal{U}(t,t_0)=\mathcal{U}_0(t,t_0)\mathcal{S}(t,t_0)$, with the free time-evolution $\mathcal{U}_0(t,t_0)$, and 
\begin{align}
\mathcal{S}(t,t_0)
&=
1 
-
i
\int_{t_0}^t dt_1 H'(t_1)
\nonumber
 \\
&
 -
\int_{t_0}^t dt_1 
\int_{t_0}^{t_1} dt_2 H'(t_1)   H'(t_2)
+
\mathcal{O}(g^3).
\label{perteh}
\end{align}
Here the time-dependence of the operators $H'$ is understood in the interaction picture, $A(t)=\mathcal{U}_0(t,t_0)^\dagger A\, \mathcal{U}_0(t,t_0)$.  The RIXS signal \eqref{gegdh} is obtained by expanding both evolution operators in $\langle b^\dagger b \rangle_t=\langle\mathcal{U}(t,t_0)^\dagger b^\dagger b\, \mathcal{U}(t,t_0)\rangle$ to second order, keeping only the orderings of the operators $P_\nu$ and $P_\nu^\dagger$ out of $H'$ which lead to a non-vanishing action on an initial state without core hole and photon. This gives
\begin{align}
 &\langle b^\dagger b \rangle_t
 =
 g^4
 \sum_{\sigma',\sigma}
\int_{t_0}^t dt_1' 
\int_{t_0}^{t_1'} dt_2'
\int_{t_0}^t dt_1 
\int_{t_0}^{t_1} dt_2 
\, s(t_2)
 s^*(t_2')
 \nonumber
 \\
&\times \,\,\Big\langle
 \mathcal{U}_0(t_0,t_2')
  P_{\text{in},\sigma'}
 e^{i\notilde \omega_{\text{in}} t_2'}
   \mathcal{U}_0(t_2',t_1')
   b
  P^\dagger_{\text{out},\sigma'}
         \mathcal{U}_0(t_1',t)
         b^\dagger
      b
 \nonumber
 \\
&\times \,\,\,\,
         \mathcal{U}_0(t,t_1)
   b^\dagger
  P_{\text{out},\sigma}
   \mathcal{U}_0(t_1,t_2)
  P_{\text{in},\sigma}^\dagger
 e^{-i\notilde \omega_{\text{in}} t_2}
 \mathcal{U}_0(t_2,t_0)
\Big\rangle_0,
 \nonumber
\end{align}
where $\langle\cdots\rangle_0$ is the expectation value in the unperturbed initial state.
For a local probe, where all operators act at the same site, the spatial factors $\eta_{\nu}$ drop out ($|\eta_{\nu}|^2=1$) and have therefore been omitted in the expression. Using the free evolution of the photon, $b(t)=e^{-i\notilde \omega_{\text{out}}(t-t_0)}$, and setting $g=1$, we have
\begin{align}
 &\langle b^\dagger b \rangle_t
 =
 \sum_{\sigma',\sigma}
\int_{t_0}^t dt_1' 
\int_{t_0}^{t_1'} dt_2'
\int_{t_0}^t dt_1 
\int_{t_0}^{t_1} dt_2 
\, s(t_2)
 s^*(t_2')
 \nonumber
 \\
&
\times 
   e^{i\notilde \omega_{\text{out}} (t_1-t_1')}
  e^{i\notilde \omega_{\text{in}} (t_2'-t_2)}
\Big\langle \mathcal{U}_0(t_0,t_2')
  P_{\text{in},\sigma'}
   \mathcal{U}_0(t_2',t_1')
  P^\dagger_{\text{out},\sigma'}
 \nonumber
 \\
&\times 
         \mathcal{U}_0(t_1',t)  \mathcal{U}_0(t,t_1)
  P_{\text{out},\sigma}
   \mathcal{U}_0(t_1,t_2)
  P_{\text{in},\sigma}^\dagger
 \mathcal{U}_0(t_2,t_0)
\Big\rangle_0.
\label{gehejkl}
\end{align}
The expectation value is now restricted to the electronic subsector. 

In equilibrium, Eq.~\eqref{gehejkl} reduces to the standard expression for the RIXS signal in terms of the absolute square of an amplitude (see appendix),
\begin{align}
\label{quasiexact}
I_{\text{RIXS}}
&=
\sum_{i,f}
w_i
|T_{i,f}|^2
\,
| \tilde s(\notilde \omega_{\text{out}}-\notilde \omega_{\text{in}}+E_f-E_i)|^2,
\\
|T_{i,f}|^2
&=
\Bigg|
\sum_{m,\sigma}
\frac{\langle \Psi_f| 
  P_{\text{out},\sigma}
   |\Psi_m\rangle
   \langle \Psi_m|
  P_{\text{in},\sigma}^\dagger
 |\Psi_i\rangle}
 {\notilde \omega_{\text{out}} +E_f-E_m+i\Gamma}
 \Bigg|^2,
 \label{gwgefgeje02}
\end{align}
where $ |\Psi_l\rangle$ and $E_l$ are initial ($l=i$), final ($l=f$), and intermediate ($l=m$) states with their respective energies,  $w_i$  is the initial state weight, and $1/\Gamma$ is the lifetime of the intermediate state. The function $\tilde s(\omega)$ is the Fourier transform of the pulse envelope. For example, for a Gaussian envelope $s(t)=e^{-t^2/2\tau^2}$ with pulse duration $\tau$, we have $|\tilde s(\omega)|^2=\pi\tau^2 e^{-(\omega\tau)^2}$. For a  long pulse duration, this becomes sharp in $\omega$, and the normalization implies $|\tilde s(\omega)|^2\to \delta(\omega)\tau$, so that one can define the standard rate
\begin{align}
\Gamma_{\text{RIXS}}
&=
\lim_{\tau\to\infty}
\frac{1}{\tau}
I_{\text{RIXS}}
\\
&=
\sum_{i,f}
w_i
|T_{i,f}|^2
\,
\delta(\notilde \omega_{\text{out}}-\notilde \omega_{\text{in}}+E_f-E_i).
 \label{gwgefgeje03}
\end{align} 
It is important to note that for long pulses the delta function in this equation indicates perfect energy conservation, which is not broadened by the lifetime of the intermediate state. Only the probe duration $\tau$ limits the energy resolution in the pulsed result \eqref{quasiexact} via the time-frequency uncertainty and the corresponding spectral width of the probe pulse. In equilibrium, the only difference between Eq.~\eqref{quasiexact} and our model is the implementation of the core-hole lifetime, which is added ad hoc in the Kramers formula, while the core-hole decay to a reservoir is included explicitly in Eq.~\eqref{gehejkl} through the system-bath Hamiltonian. We will demonstrate, however, that the choice of the bath \eqref{cbath}  is basically equivalent to a Lorentzian broadening as  in Eq.~\eqref{gwgefgeje02}  (see Sec.~\ref{sec:bench:0}).

\begin{figure}[tbp]
\centerline{\includegraphics[width=0.4\textwidth]{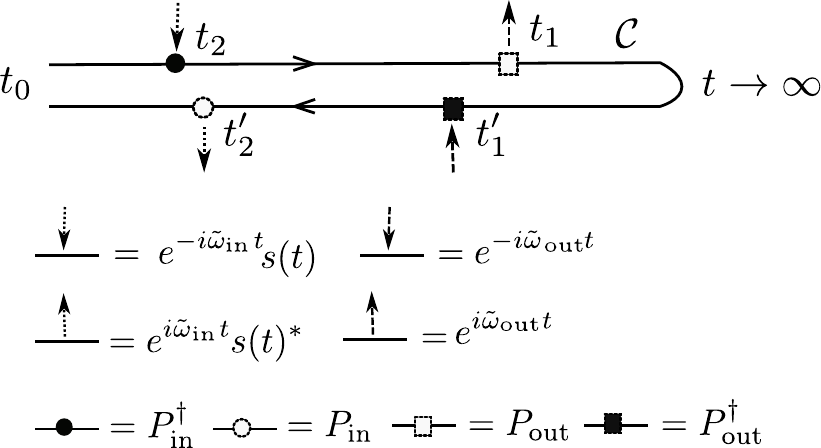}}
\caption{
Diagrammatic representation of Eq.~\eqref{gehejkl}. Arrows along the contour $\mathcal{C}$ from time $t$ to $t'$ denote the time evolution, further symbols are  explained in the bottom lines (operators $P_{\nu}$  and $P_{\nu}^\dagger$, as well as ingoing and outgoing lines for the phase factors $e^{\pm i\omega_{\nu} t}$ and probe envelope corresponding to photon annihilation and creation). The RIXS signal is obtained by extending the contour from a time $t_0$ before the RIXS pulse to $t=\infty$, and integrating over the internal times by keeping their order on the contour. 
}
\label{fig00}
\end{figure}

To proceed with the evaluation of the time-dependent result \eqref{gehejkl}, it will be convenient to represent the equation 
in terms of a contour ordered expectation value on the Keldysh contour. In general, the time-dependent expectation value of an operator $A$ can be denoted as the contour-ordered expectation value with the action $\mathcal{S}$
\begin{align}
\label{expa}
\langle
A(t)
 \rangle_{\mathcal{S}}
 \equiv
\frac{1}{Z}
\text{tr}
\big[T_\mathcal{C}e^{\mathcal{S}}
A(t)\big],
\end{align}
where $T_\mathcal{C}$ orders operators along the Keldysh contour that evolves first forward in time along an upper branch and then backward in time along a lower branch. Similarly, we introduce higher order contour-ordered correlation functions
$\langle
A(t)B(t')
\cdots
 \rangle_{\mathcal{S}}$, and we will use the convention that  $t_{\pm}$ denotes a real time argument on the upper ($+$) or lower ($-$) branch. The expectation value in the integral in Eq.~\eqref{gehejkl}  is therefore understood as a contour-ordered expectation value 
\begin{align}
 \Big\langle 
P_{\text{in},\sigma'}(t_{2,-}')
  P^\dagger_{\text{out},\sigma'}(t_{1,-}')
  P_{\text{out},\sigma}(t_{1,+})
  P_{\text{in},\sigma}^\dagger(t_{2,+})
\Big\rangle_{\mathcal{S}_{0}} 
\end{align}
with the unperturbed impurity action $\mathcal{S}_{0}$, i.e, Eq.~\eqref{simp} evaluated at $g=0$. The integral \eqref{gehejkl} is thereby represented by an intuitive diagrammatic representation, analogous to double-sided Feynman diagrams in nonlinear optics, see Fig.~\ref{fig00}.

\subsection{Second order perturbation theory (\CMethod{}  approach)}
\label{sec:forma:O2}

Instead of doing the perturbation theory in both the coupling to ingoing and outgoing modes, one could also leave the driving field $H_{\text{in}}$  [Eq.~\eqref{ingoung}] with an amplitude $g_{\text{in}}\neq0$ explicitly in the action, evaluate the RIXS signal $I_{\text{RIXS}}(g_{\text{in}})$ to second order in the coupling to the outgoing mode, and extract the leading second order contribution in $g_{\text{in}}$ from the numerical data.  Using the perturbative expansion $\eqref{perteh}$ to first order, with only $\notilde H_{\text{out}}$ [Eq.~\eqref{outgoung}] as perturbation, we have
\begin{align}
 \langle b^\dagger b \rangle_t
 =&
 g_\text{out}^2
 \sum_{\sigma',\sigma}
\int_{t_0}^t dt_1' \,dt_1 
\Big\langle
   \mathcal{U}_{\text{in}}(t_0,t_1')
   b
  P^\dagger_{\text{out},\sigma'}
         \mathcal{U}_{\text{in}}(t_1',t)
 \nonumber
 \\
&\hspace{5mm}\times \,\,
         b^\dagger
      b\,
         \mathcal{U}_{\text{in}}(t,t_1)
   b^\dagger
  P_{\text{out},\sigma}
   \mathcal{U}_{\text{in}}(t_1,t_0)
\Big\rangle_{0},
\label{gehejkl04}
\end{align}
where the subscript ``$\text{in}$" of the evolution operator indicates that the time evolution $ \mathcal{U}_{\text{in}}(t_1,t_0)$ includes the driving field $H_{\text{in}}$. The equivalence of Eqs.~\eqref{gehejkl04} and \eqref{gehejkl} can be seen by expanding $ \mathcal{U}_{\text{in}}$ in Eq.~\eqref{gehejkl04} to first order in the perturbation $H_{\text{in}}$, using the 
expansion \eqref{perteh}. Again factoring out the photon operators 
in Eq.~\eqref{gehejkl04}, and setting  $g_\text{out}=1$, we have
\begin{align}
\label{o2}
 &I_{\text{RIXS}}(g_{\text{in}})= 
 \sum_{\sigma',\sigma}
\int dt_1' \,dt_1 
\,e^{i\omega_{\text{out}}(t_1-t_1')}
\mathcal{P}_{\sigma',\sigma}(t_{1-}',t_{1,+}), 
\end{align}
where  we have introduced the contour-ordered expectation value 
\begin{align}
\label{PP}
\mathcal{P}_{\sigma',\sigma}(t',t) 
=
\big\langle
  P^\dagger_{\text{out},\sigma'}(t')
  P_{\text{out},\sigma}(t)
\big\rangle_{\mathcal{S}_\text{in}},
\end{align}
evaluated with the action $\mathcal{S}_\text{in}$ of the driven model.
The RIXS amplitude may thus be obtained by evaluating  $\mathcal{P}_{\sigma',\sigma}(t'_{-},t_{+}) $
in the driven model for small $g_{\text{in}}$, and extracting the limit 
\begin{align}
 &I_{\text{RIXS}}=
\lim_{g_{\text{in}}\to 0} g_{\text{in}}^{-2} I_{\text{RIXS}}(g_{\text{in}})
\end{align}
numerically. It should be mentioned that the integration range is effectively restricted: The action of the operators $P$ gives zero for times $t,t'$ before the probe pulse and after the decay of the core-hole, i.e., for times much later that $\Gamma^{-1}$ after the probe. Hence the correlation function $\mathcal{P}$ [Eq.~\eqref{PP}] has to be evaluated only in a small time window.

\subsection{Direct evaluation (\CFMethod{}  approach)}
\label{sec:forma:O0}

To avoid the calculation of the second-order correlation function \eqref{PP}, one can simply include both the driving term and the outgoing photon state into an explicit time-dependent simulation with an amplitude $g\neq 0$, evaluate the expectation value 
$N(g,t)=\langle b^\dagger(t)b(t)\rangle_{\mathcal{S}_\text{imp}}$ from the full action \eqref{simp}, and extract the limit 
\begin{align}
\label{0thoder}
I_{\text{RIXS}}=\lim_{t\to\infty} \lim_{g\to0}  g^{-4} N(g,t)
\end{align}
numerically. The order of the limits is important, as taking the limit $t\to\infty$ first will lead to a photon occupation of order one on times which diverge with $1/g$. In practice, the $t\to\infty$ limit again saturates within a short time determined by the core-hole lifetime, so that no long-time simulations are needed.

The direct evaluation \eqref{0thoder} is particularly suitable if the model is solved within an expansion in the system-bath coupling, i.e, the hybridization expansion. The latter can be formulated for any local Hilbert space and local Hamiltonian. The numerical effort increases exponentially with the local Hilbert space, and simulations including explicitly a bosonic Hilbert space can be demanding.\cite{Grandi2020} However, in the present case one can restrict the local Hilbert space from the outset to only those states which contribute in leading order to the RIXS signal. If $\{ |\alpha\rangle, \alpha=1,..., d\}$ is  a basis for the valance band manifold, $\{ |0\rangle, |\!\uparrow\rangle, |\!\downarrow\rangle , |\!\uparrow\downarrow\rangle\}$ the basis for the core orbitals, and $\{ |n_\text{ph}\rangle,n=0,1,2,...\}$ the basis for the photon, the RIXS signal can be computed after projecting the impurity model to the subspace 
\begin{align}
\{
|\alpha,\uparrow\downarrow,0_\text{ph}\rangle,\,
|\alpha,\sigma,0_\text{ph}\rangle,\,
|\alpha,\uparrow\downarrow,1_\text{ph}\rangle,\,
\alpha=1,...,d; \sigma=\downarrow,\uparrow
\}.
\nonumber
\end{align}
The relevant interaction operators $P^\dagger b$ and $Pb^\dagger$ act within this manifold. Higher order excited states such as $|\alpha,\uparrow\downarrow,2_\text{ph}\rangle$
do not appear as initial, final, or intermediate states in the time evolution for the perturbative expression \eqref{gehejkl}, and can therefore be omitted in the time evolution of the model from the outset. 

In the \CMethod{}  approach, the simulation of the driven model can be restricted to the electronic subsector, spanned by the states 
\begin{align}
\{
|\alpha,\uparrow\downarrow,0_\text{ph}\rangle,\,
|\alpha,\sigma,0_\text{ph}\rangle,\,
\alpha=1,...,d; \sigma=\downarrow,\uparrow
\}.
\nonumber
\end{align}
Hence the dimensionality of the relevant Hilbert space is smaller in the \CMethod{}  method, but the difference in the numerical effort is moderate. The main additional numerical cost of the direct \CFMethod{}   approach as compared to the \CMethod{}  approach lies in the number of independent simulations that are required. In the \CMethod{}  approach one has to perform a separate simulation for each probe frequency $\omega_{\text{in}}$ (since the probe is included explicitly). The results for each $\omega_{\text{out}}$ can then  be obtained by a straightforward convolution \eqref{o2}. In the \CFMethod{}  approach, in contrast, a separate time-dependent solution of the impurity model is required for each ingoing and outgoing frequency.

\subsection{Vertex corrections}
\label{sec:forma:nca}

\begin{figure}[tbp]
\centerline{\includegraphics[width=0.5\textwidth]{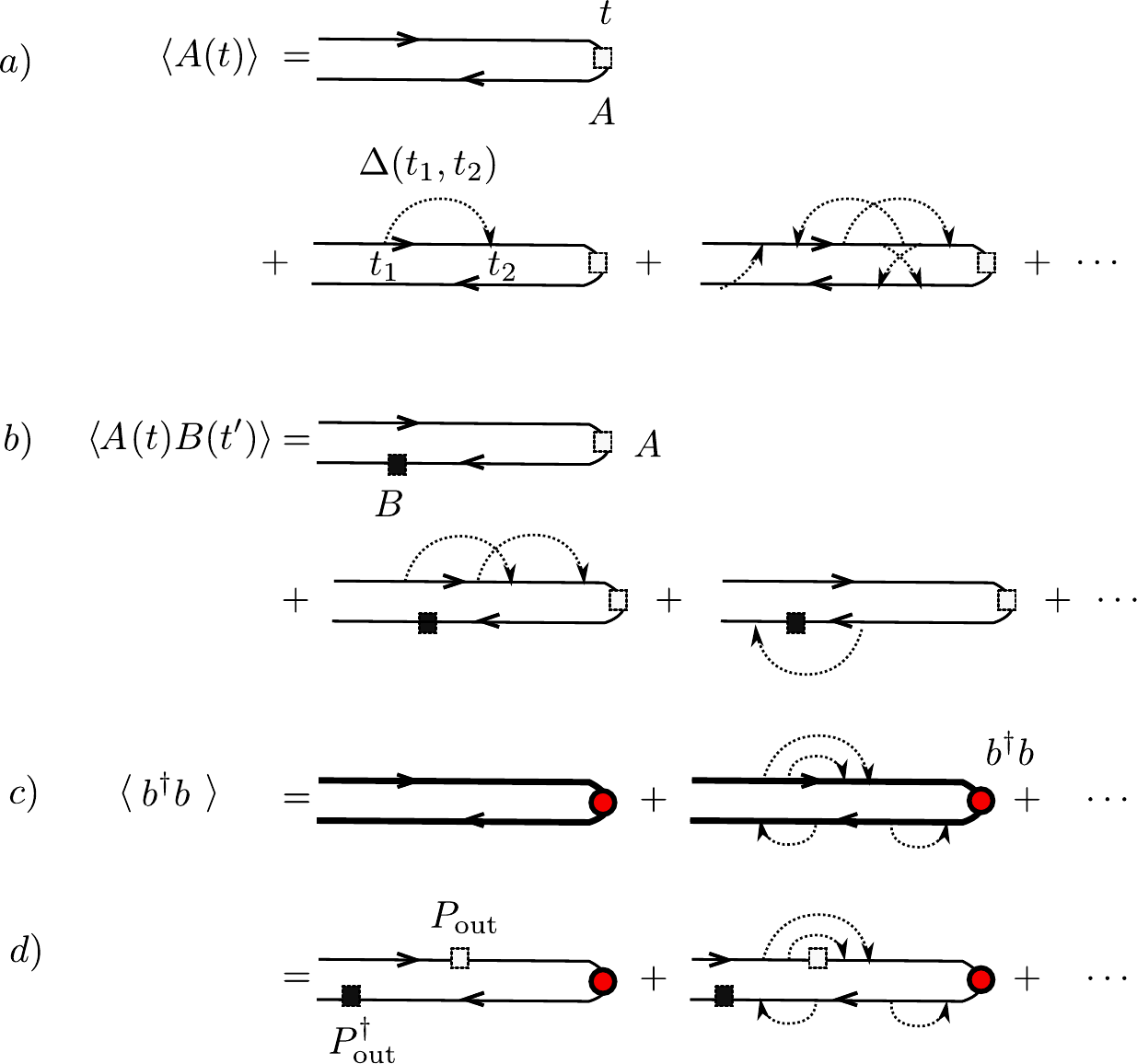}}
\caption{
Illustration of hybridization events (description in the main text).
}
\label{fig02}
\end{figure}

A rather versatile approach to solving the impurity model \eqref{simp} is the systematic expansion in the hybridization function. Out of equilibrium, this can be done perturbatively order by order,\cite{Eckstein2010nca} or using hybridization-expansion Quantum Monte Carlo simulations.\cite{Muehbacher2008, Werner2009, Gull2010,Cohen2015} Both within the perturbative framework, and within Quantum Monte Carlo schemes that are based on re-summations of diagrams,\cite{Gull2010, Cohen2015} the evaluation of high-order correlation functions is challenging because of vertex corrections:
The hybridization expansion expands the time evolution operator in terms of hybridization events as illustrated  diagrammatically in Fig.~\ref{fig02}. A dashed arrow from time $t$ to $t'$ corresponds to the emission of an electron from the system into the bath at time $t$ and the absorption of an electron at a later time $t'$, i.e, the hybridization function $\Delta$. In order to obtain the full evolution one must sum over all such events on the time contour, where the bare time evolution along the contour corresponds to the isolated impurity ($H_\text{loc}$). For all hybridization expansion schemes based on a re-summation of such diagrams, the most straightforward step is to sum all diagrams in the time-evolution operator, and in the expression for the expectation value $\langle A(t)\rangle$  of some operator $A$ at the end of the contour (see Fig.~\ref{fig02}a). However, for a two-point correlation function of two operators $A$ and $B$ one needs to include also terms which connect the time-evolution operator before and after $B$ via a hybridization line, i.e., events in which an  electron is emitted into the reservoir {\em before} the action of $B$ and absorbed {\em after} that (see third term in Fig.~\ref{fig02}b). The systematic re-summation of such vertex corrections for out-of-equilibrium Green's functions is typically challenging.

In the present work, we will exemplarily solve some models within the leading-order hybridization expansion, which sums up diagrams with non-crossing hybridization lines.\cite{Eckstein2010nca} 
 In the \CFMethod{} approach one would  evaluate the expectation value of the operator $b^\dagger b$ at the end of the contour, including all non-crossing hybridization lines (Fig.~\ref{fig02}c). Here, the bare evolution includes the light-matter coupling and is illustrated by thick lines. Extracting the leading time-dependent perturbation in the dipolar operator from this result  corresponds to replacing the bare evolution by the evolution at zero dipolar coupling, and inserting operators $P$ and $P^\dagger$ at the corresponding times  (Fig.~\ref{fig02}d). As the diagrams illustrate, this automatically generates a class of ladder-type vertex corrections, such as the second diagram in (Fig.~\ref{fig02}d), which are not included in the NCA evaluation of the  correlation function $\langle P^\dagger(t)P(t')\rangle$ that enters the  \CMethod{} approach.  In this sense, the \CFMethod{} approach is more accurate. The  fact that the direct evaluation of the expectation value $\langle b^\dagger b\rangle$ within the \CFMethod{} framework implicitly contains a large class of diagrams that is not easily reproduced in a diagrammatic representation of the response function is common to general time-dependent diagrammatic theories, and will also hold for higher order variants of the hybridization expansion, as well as for typical weak-coupling expansions  or Quantum Monte Carlo schemes that are based on a re-summation of diagrams.

\section{Results}
\label{sec:bench}

In this section, we evaluate and compare the formalisms introduced in the previous section for several problems of increasing complexity: An impurity model without any bath (atomic limit), an impurity model with a single bath orbital, a single-impurity Anderson model, and a non-equilibrium DMFT simulation for the dynamics of a Mott antiferromagnet. In all cases, the time-dependent impurity model will be solved within the non-crossing approximation, see the discussion in Sec.~\ref{sec:forma:nca}. We use a Gaussian probe pulse
\begin{align}
\label{rixsspulse}
s(t)=0.1e^{-(t-t_p)^2/\tau^2}
\end{align}
in Eq.~\eqref{field}, with a probe duration $\tau$, centered around a probe time $t_p$. 

\subsection{Single-impurity model in the atomic limit}
\label{sec:bench:0}

The simplest test model contains just a single spinful valence orbital and no hybridization term \eqref{sdd}. The valence Hamiltonian $H_{dd}$ in Eq.~\eqref{hloc_RWA} is then given by 
\begin{align}
\label{hdot01}
H_{dd} = \epsilon \big(c_{\uparrow}^\dagger c_{\uparrow} + c_{\downarrow}^\dagger c_{\downarrow}\big)
+ U c_{\uparrow}^\dagger c_{\uparrow}  c_{\downarrow}^\dagger c_{\downarrow},
\end{align}
with an on-site energy $\epsilon$ and a local Hubbard interaction $U$. Because there is no $d$-level hybridization function $\Delta$, there are no vertex corrections in the evaluation of the correlation function $\mathcal{P}$ [Eq.~\eqref{PP}] within the NCA framework, and the \CMethod{}  and \CFMethod{}  approaches are therefore equivalent by construction. The only approximation which enters the numerical solution of this model within the present framework is the treatment of the core bath within the NCA approximation.
 We therefore compare the solution of this model with the exact-diagonalization result \eqref{quasiexact} with a level broadening $\Gamma$, in order to confirm that the core bath setup as defined in Eqs.~\eqref{bathcc} and \eqref{cbath} provides an accurate description of an exponential core hole decay with time constant $1/\Gamma$.
 
With the Hamiltonian Eq.~\eqref{hdot01}, the following processes $|\Psi_i\rangle\to |\Psi_m\rangle\to |\Psi_f\rangle$  appear in the sum \eqref{gwgefgeje02} for the exact diagonalization result:
\begin{align}
&|\sigma\rangle \to |\!\uparrow\downarrow\rangle  \to |\sigma\rangle,
\,\,\,E_i=\epsilon, \,\,\,E_m=2\epsilon+U-2U_{cd},
\label{path1}
\\
&|0\rangle \to |\sigma\rangle \to |0\rangle,
\,\,\,E_i=0, \,\,\,E_m=\epsilon-U_{cd}.
\label{path2}
\end{align}
(Only the valence orbital configuration is shown, not the core-hole in  the intermediate state.) For the single-site model, the initial and final states are always the same ($E_i=E_f$). The thermal occupations of the initial state are given by
\begin{align} 
w_0&=z^{-1},\,\,\,\,
w_\sigma=z^{-1} e^{-\beta\epsilon},\,\,\,\,
\end{align}
with $z=1+2e^{-\beta\epsilon} + e^{-\beta(U+2\epsilon)}$, and the matrix elements $\langle \Psi_f|   P_{\text{out},\sigma}   |\Psi_m\rangle   \langle \Psi_m|  P_{\text{in},\sigma}^\dagger |\Psi_i\rangle$ for the transition are unity. Hence, Eq.~\eqref{quasiexact} gives
\begin{align}
&I_{\text{RIXS}}
 =
\frac{\pi}{\Gamma}
\Big[
2 w_\sigma
L_\Gamma\big(\notilde \omega_{\text{out}} -(\epsilon+U-2U_{cd})\big)
\nonumber \\&\quad\quad+
4w_0
L_\Gamma\big(\notilde \omega_{\text{out}} -(\epsilon-U_{cd})\big)
 \Big]
\big| \notilde s(\notilde \omega_{\text{out}}-\notilde \omega_{\text{in}})\big|^2,
 \label{gwgefgeje08}
\end{align}
where $L_\Gamma(x) = \frac{\Gamma/\pi}{x^2+\Gamma^2}$ is a normalized Lorentzian peak. In passing we remark that the weight of the term with initial state $|0\rangle$ is $4w_0$, while it is just $2w_\sigma$ in total for the contribution of both initially singly occupied states. This is due to an interference of the two paths with the same initial state but different intermediate states in \eqref{path2}, while the two contributions in \eqref{path1} differ only in the initial state and are just added up. One therefore cannot simply calculate the RIXS signal from a spin-polarized level, but in general both spin contributions must be included in the numerical solution of the time-dependent impurity model.

\begin{figure}[tbp]
\centerline{\includegraphics[width=0.5\textwidth]{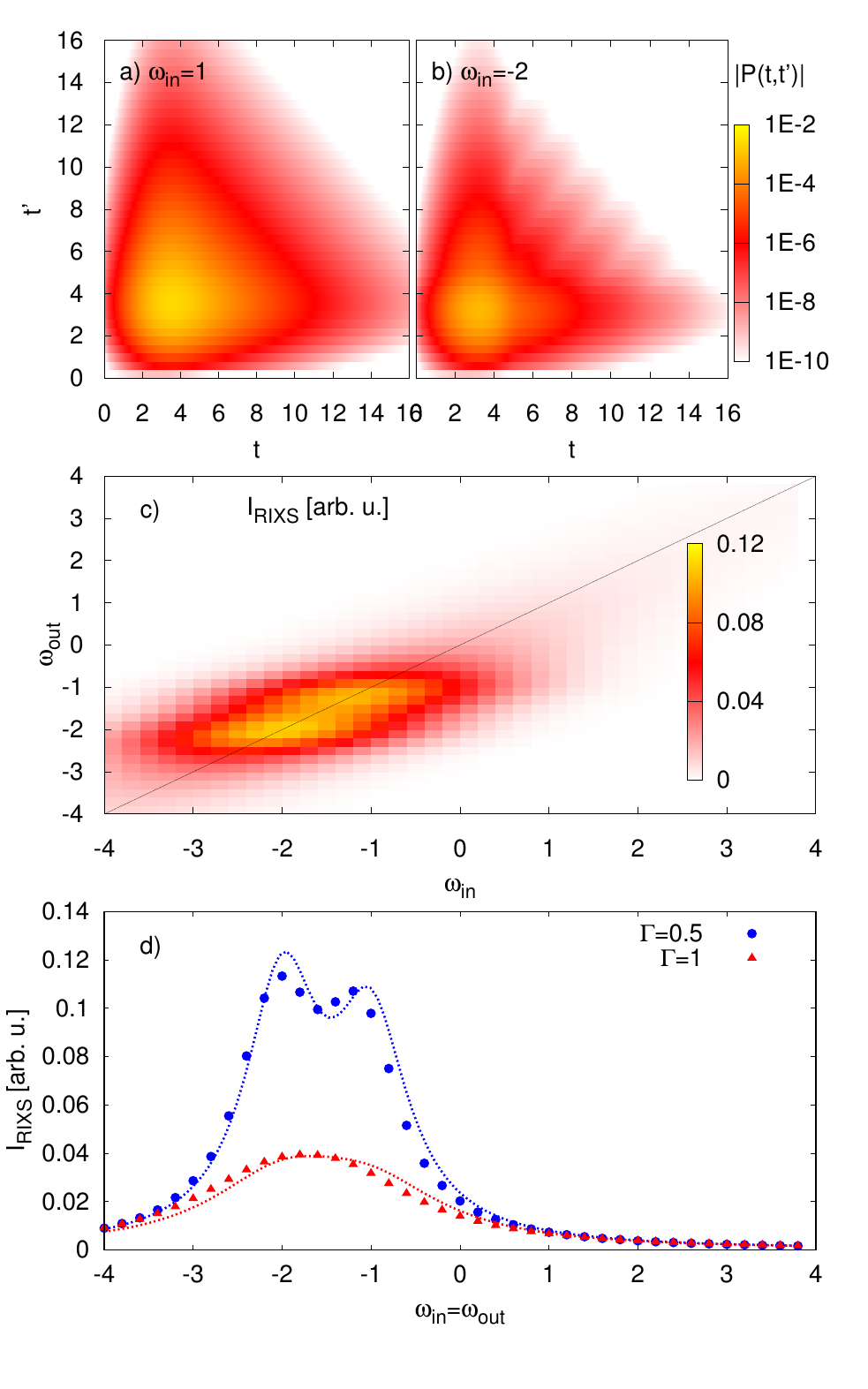}}
\caption{
a) and b) The two-point correlation function $\mathcal{P}_{\uparrow\uparrow}(t,t')$ for the single-orbital model \eqref{hdot01} at inverse temperature $\beta=5$, for $U=2$, $\epsilon=-U/2$, $U_{cd}=0$, core-bath coupling $\Gamma=1$ and $\omega_{\text{in}}$ as indicated. The RIXS pulse is given by Eq.~\eqref{rixsspulse} with duration $\tau=2$ and probe time $t_p=4$. c) The spectrum $I_{\text{RIXS}}(\omega_{\text{in}},\omega_{\text{out}})$ for the same parameters. d) A cut of $I_{\text{RIXS}}(\omega_{\text{in}},\omega_{\text{out}})$ along the elastic line $\omega_{\text{in}}=\omega_{\text{out}}$, compared to the exact diagonalization result \eqref{quasiexact}
with a level broadening $\Gamma$, for two different values of $\Gamma$.}
\label{fig03}
\end{figure}

Figure~\ref{fig03}  exemplarily shows the function $\mathcal{P}_{\uparrow\uparrow}(t,t')$  [Eq.~\eqref{PP}] for the single impurity model in the atomic limit at inverse temperature $\beta=5$, for $U=2$, $\epsilon=-U/2$, $U_{cd}=0$, $\Gamma=1$, and $\omega_{\text{in}}$ as indicated. The core-hole decay time $1/\Gamma$ is comparable to other atomic time scales, as is typical for realistic systems. Figure~\ref{fig03}a correspond to a frequency $\omega_{\text{in}}=1$ which is resonant to the doublon creation energy $\notilde \omega_{\text{in}}=U+2\epsilon-\epsilon=U+\epsilon=U/2$, while the frequency $\omega_{\text{in}}=-2$ (Fig.~\ref{fig03}b) is off-resonant. In both cases the explicit incorporation of the bath \eqref{bathcc} leads to a decay of the  correlation function $\mathcal{P}(t,t')$ after the pulse, i.e., for $(t'-t_p),(t-t_p)\gg 1/\Gamma$. This decay of the two-point correlation function due to the core lifetime is more or less independent of the valence Hamiltonian and therefore holds also for the more involved examples below. It allows to cut off the time-dependent simulations at times $t-t_p\gg 1/\Gamma$. 

Figure \ref{fig03}c shows the RIXS spectrum obtained from the integral \eqref{o2}. Because there is only an elastic process in the single-orbital model, the spectrum $I_{\text{RIXS}}(\omega_{\text{in}},\omega_{\text{out}})$ consists of a singe line at $\omega_{\text{in}}=\omega_{\text{out}}$, broadened by the energy uncertainty of the incoming pulse. This is consistent with the analytic result Eq.~\eqref{gwgefgeje08}.  A comparison of the spectra as a function of $\omega_{\text{in}}$ along the elastic line $\omega_{\text{in}}=\omega_{\text{out}}$
with Eq.~\eqref{gwgefgeje08} demonstrates a good quantitative agreement for different core-hole lifetimes. (The larger lifetime $\Gamma=1/2$ resolves the difference between the two possible intermediate states, while the shorter lifetime yields only one broad peak.) This confirms that the explicit treatment of the core-hole bath in our formalism is consistent with the level broadening
in the conventional exact diagonalization formula \eqref{quasiexact}. 

\subsection{Impurity model with one bath site}
\label{sec:bench:one}

Next we apply the formalism to an impurity model with a single bath orbital. The valence Hamiltonian is given by
\begin{align}
H_{dd}
&=
H_{d,\text{at}} + J \sum_{\sigma} (c_{\sigma}^\dagger a_{\sigma} + h.c.) +\epsilon_\text{bath}\sum_{\sigma}  a_{\sigma} ^\dagger a_{\sigma},
\label{Honebath}
\end{align}
where the atomic Hamiltonian $H_{d,\text{at}}$ is the same as Eq.~\eqref{hdot01}, $a_{\sigma}$ ($a_{\sigma}^\dagger$) is the annihilation (creation) operator of an electron with spin $\sigma$ in the bath orbital, and $J$ the tunnelling matrix element. This corresponds to a hybridization function 
\begin{align}
\Delta(t,t') = J^2 g(t,t'),
\end{align}
in the action \eqref{sdd}, where $g(t,t')$ is the propagator of the isolated bath site at energy $\epsilon_\text{bath}$. We consider the particle-hole symmetric set-up $U=2$, $\epsilon=-U/2$, $\epsilon_\text{bath}=0$.  For an impurity solver based on the expansion in the hybridiziation $\Delta(t,t')$, the two-site model \eqref{Honebath} is not necessarily faster convergent than a model with a continuous bath, while the result for the RIXS signal can still be compared to the analytic relation \eqref{quasiexact}. Though 
still rather simple in terms of physics, the model \eqref{Honebath} therefore provides a nontrivial test case to illustrate the two formalisms for evaluating the RIXS signal.

\begin{figure}[tbp]
\centerline{\includegraphics[width=0.45\textwidth]{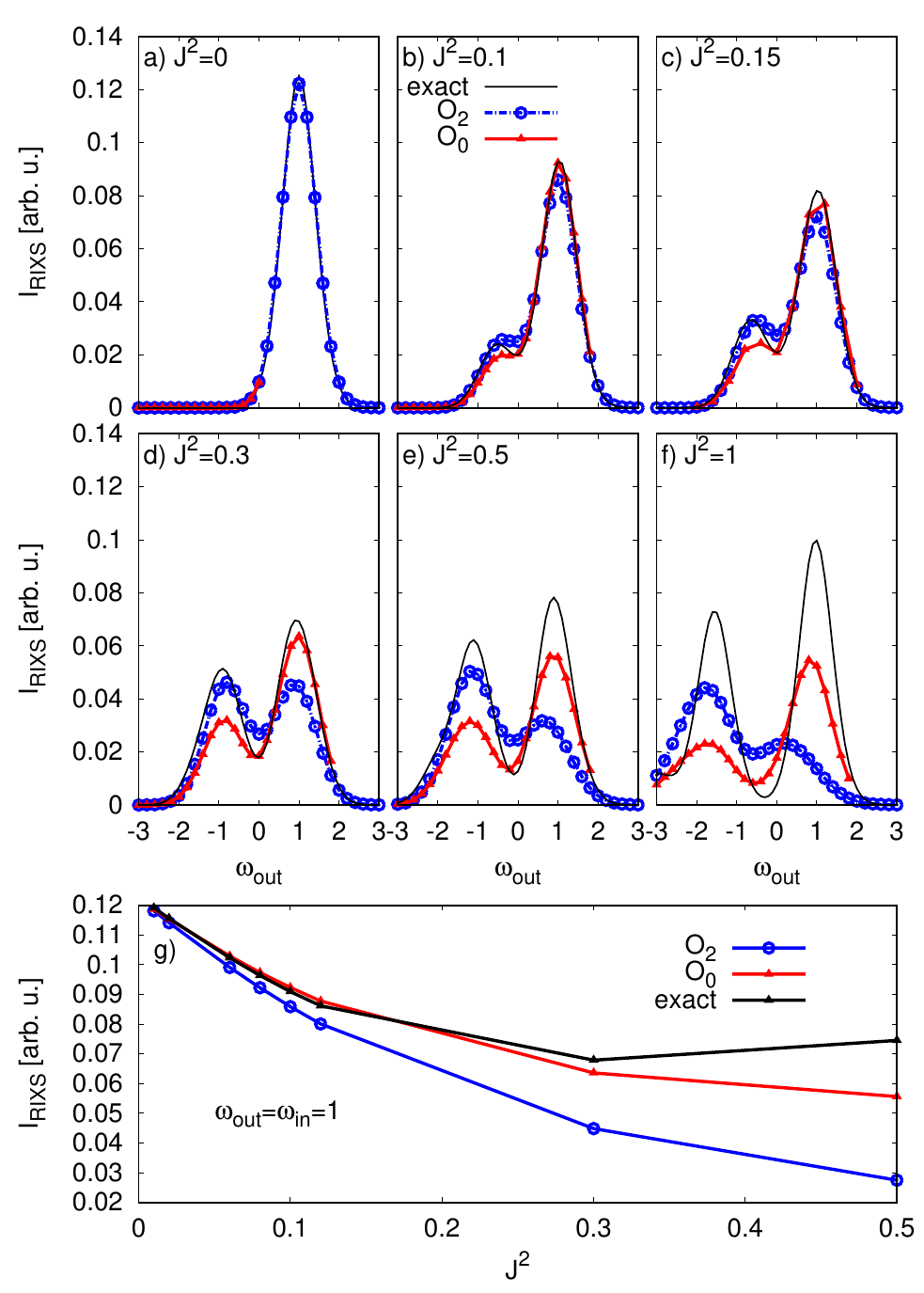}}
\caption{
a)-f)  $I_{\text{RIXS}}(\omega_{\text{in}},\omega_{\text{out}})$ for the impurity model \eqref{Honebath} with one bath orbital for various tunnelling matrix elements $J$ (inverse temperature $\beta=5$, $U=2$, $\epsilon=-U/2$, $U_{cd}=0$, bath coupling $\Gamma=1$, and $\omega_{\text{in}}=1$, and a  RIXS probe pulse \eqref{rixsspulse} with duration $\tau=2$). The black line is from exact diagonalization [Eq.~\eqref{quasiexact}], while circles and triangles correspond to the results \eqref{o2} and \eqref{0thoder} obtained from the \CMethod{}  and \CFMethod{}  approaches, respectively (see legend in panel b). g) Comparison of the signal $I_{\text{RIXS}}(\omega_{\text{in}},\omega_{\text{out}})$ on the elastic line ($\omega_{\text{in}}=\omega_{\text{out}}=1)$ as a function of $J^2$ for the three approaches.}
\label{fig04}
\end{figure}

Figure \ref{fig04} compares the exact RIXS spectrum [Eq.~\eqref{quasiexact} applied to model \eqref{Honebath}] with the results \eqref{o2} and \eqref{0thoder} obtained from the \CMethod{}  and \CFMethod{}  approaches, respectively. The pulse parameters are the same as in Fig.~\ref{fig03}, with pulse duration $\tau=2$ and  damping $\Gamma=1$. We analyze the spectrum as a function of $\notilde \omega_{\text{out}}$, where $\notilde \omega_{\text{in}}=1$ is fixed to be resonant to the intermediate doublon state. (Because of the large damping $\Gamma$, the results are qualitatively similar for different $\notilde \omega_{\text{in}}$.) For hopping $J=0$, there is only the elastic  line at $\notilde \omega_{\text{out}}=\notilde \omega_{\text{in}}$, broadened by the frequency uncertainty of the Gaussian probe (Fig.~\ref{fig04}a). With increasing tunnelling $J$, a second peak (the loss peak) appears around  $\notilde \omega_{\text{out}}=\omega_{\text{in}}-\Delta_U$, corresponding to processes in which the system is left with a doublon (doubly occupied site) and a hole after the RIXS process. The  corresponding excitation energy is $\Delta_U=U$, with corrections of order $J^2/U$ for small $J$. 

Both the \CMethod{}  and \CFMethod{}  approach are basically exact by construction for $J=0$. For $J>0$, the two approaches differ because vertex corrections are missing in the NCA solution of the two-point correlation function $\mathcal{P}(t,t')$, see Sec.~\ref{sec:forma:nca}. For larger $J$, the \CMethod{}   approach apparently describes the loss peak  better than the \CFMethod{}  approach, while the \CFMethod{} approach is better at the elastic peak. Within the \CMethod{}  approach, the loss peak becomes eventually stronger than the elastic peak (Figs.~\ref{fig04}e and f), which is reversed by the vertex corrections introduced in the \CFMethod{} method.  A quantitative comparison of the signal  $I_{\text{RIXS}}(\notilde \omega_{\text{out}},\notilde \omega_{\text{in}}) $ for the elastic peak $\notilde \omega_{\text{out}}=\notilde \omega_{\text{in}}$ shows that the  \CFMethod{}  approach is correct to leading order in $J^2$, while the \CMethod{}  deviates (Fig.~\ref{fig04}g). This is expected because vertex corrections to $\mathcal{P}(t,t')$ can arise already from a single hybridization event (see the third diagram in Fig.~\ref{fig02}b),
so that neglecting them can imply an error at the leading order $J^2$.

While both methods are not particularly accurate for large $J$, due to the NCA approximation, the results illustrate how the comparison of the two approaches can be used to estimate the importance of vertex corrections in the evaluation of the $\mathcal{P}$ correlation function, and hence for the RIXS signal. In general, one could use the numerically more costly direct  \CFMethod{}  approach to validate the quality of the \CMethod{}  approach for certain parameters, or take the deviation between the two methods as a heuristic measure for the overall error. The same argument can be used for different impurity solvers, such as higher order variants of the strong-coupling expansion, weak-coupling based solvers, or potentially real-time quantum Monte Carlo.

\begin{figure}[tbp]
\centerline{
\includegraphics[width=0.5\textwidth]{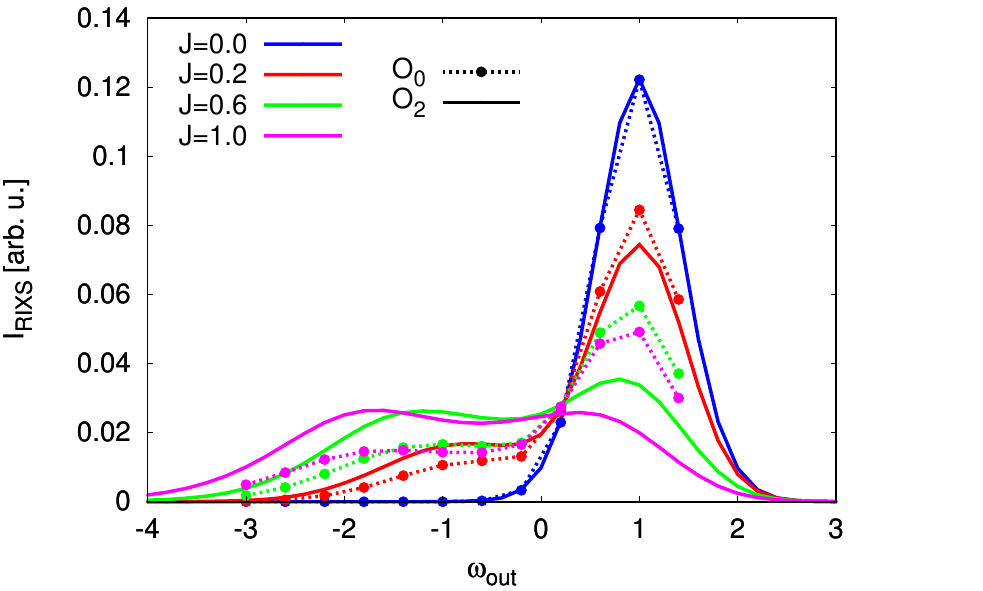}}
\caption{
a) $I_{\text{RIXS}}(\omega_{\text{in}},\omega_{\text{out}})$ for a single impurity Anderson model with hybridization function \eqref{bethe} and different hybridization strengths $J$ as indicated (inverse temperature $\beta=5$, $U=2$, $\epsilon=-U/2$, $U_{cf}=0$, bath coupling $\Gamma=1$, and $\omega_{\text{in}}=1$, and a  RIXS probe pulse \eqref{rixsspulse} with duration $\tau=2$). Circles and dashed lines correspond to the RIXS signal evaluated with the \CFMethod{}  approach [Eq.~\eqref{0thoder}], solid lines to the \CMethod{}  approach [Eq.~\eqref{o2}].}
\label{fig05}
\end{figure}

\subsection{Anderson Impurity model}
\label{sec:bench:siam}

To further illustrate the approach we consider a model where an analytic result \eqref{quasiexact} is not available. We focus on the single impurity Anderson model, i.e., model \eqref{simp} in which the local Hamiltonian $H_{dd}$ is given by a single orbital [Eq.~\eqref{hdot01}], and the hybridization function $\Delta$ in \eqref{sdd} is chosen to have a semi-elliptic density of states
\begin{align} 
\label{bethe}
-\frac{1}{\pi}\text{Im}\Delta^{\text{ret}}(\omega+i0)
=
\frac{2J^2}{\pi W} \sqrt{1-(\omega/W)^2}
\end{align}
with half bandwidth $W=2$ and overall hybridization strength $J^2$. Figure~\ref{fig05} shows the RIXS signal, evaluated with the same pulse parameters as in Fig.~\ref{fig04}. Again one can see the emergence of a loss feature at $\omega_{\text{out}}<\omega_{\text{in}}$, which is now  broadened not only due to the energy uncertainty of the pulse, but because there is an excitation continuum in the impurity model. The comparison of the two approaches gives confidence that this loss feature is qualitatively and even quantitatively captured up to relatively large hybridization strength. 

\subsection{Dynamics in the antiferromagnetic Mott insulator}
\label{sec:bench:afm}

Finally, we present an application of the formalism to a  non-equilibrium problem. We study the single-band Hubbard model
\begin{align}
H
&=
U\sum_{j}n_{j\uparrow}n_{j\downarrow}
-
 J \sum_{\sigma \langle ij\rangle}c_{i\sigma}^\dagger c_{j\sigma}
 \label{Hubbard}
\end{align}
at half filling on a bipartite lattice. Here $c_{j\sigma}$ is the fermion operator for an electron with spin $\sigma $ on site $j$ of the lattice, $U$ is the on-site interaction, and $J$ the hopping between nearest neighbors. We focus on the Mott insulating phase, with $U\gg J$. On a bipartite lattice, the system is antiferromagnetically ordered at low temperature. Here we study the system out of equilibrium, in a situation in which the antiferromagnetic order evolves on ultra-fast timescales. The RIXS signal is sensitive to the local environment of a site on the lattice, and can therefore potentially be used to monitor the evolution of the antiferromagnetic order. The reason is that in a N\'eel-ordered state at a site next to one with spin $\sigma$ is predominantly occupied with the opposite spin flavor. This enhances the possibility of charge fluctuations between neighboring sites, and thus the probability that a charge excitation (a doubly-occupied and an empty site) is generated by the RIXS process. We thus expect that the magnitude of the loss peak around $\omega_{\text{out}}\approx\omega_{\text{in}}-U$ increases with the antiferromagnetic order. 

\begin{figure}[tbp]
\centerline{
\includegraphics[width=0.5\textwidth]{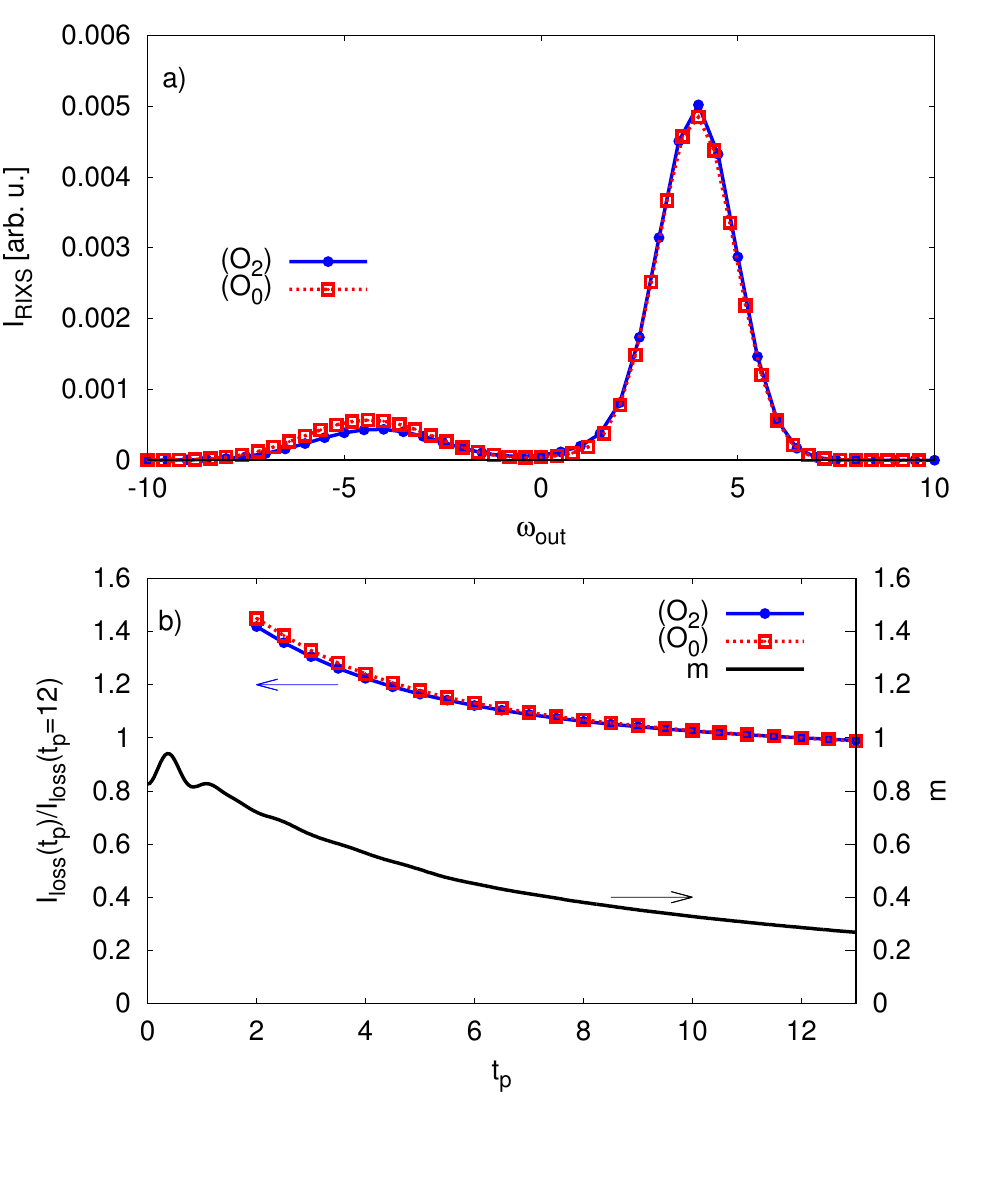}}
\caption{
a) $I_{\text{RIXS}}(\omega_{\text{in}},\omega_{\text{out}})$ for the Hubbard model at $U=8$ in the antiferromagnetic Mott phase, as described in the main text (inverse temperature $\beta=10$, $U=8$). The RIXS signal is obtained for bath coupling $\Gamma=1$, $\omega_{\text{in}}=4$, and a Gaussian probe pulse with duration $\tau=1$ and $t_p=7$. Circles and squares show results from the  \CMethod{}  approach [Eq.~\eqref{o2}] and the \CFMethod{}  approach [Eq.~\eqref{0thoder}], respectively. b) Order parameter $m$ as a function of time (solid black line, right vertical axis) after an interaction quench, and amplitude of the loss peak $I_\text{loss}(t_p)=I_{\text{RIXS}}(\omega_{\text{in}},\omega_{\text{out}})$ at $\omega_{\text{out}}=\omega_{\text{in}}-U$ in the \CMethod{}  and \CFMethod{}  approach.}
\label{fig06}
\end{figure}

To confirm this, we compute the RIXS signal for a particular situation, in which the evolution of the antiferromagnetic order itself has already been studied in detail in Ref.~\onlinecite{Werner2012}. The Hubbard model is solved using DMFT on a Bethe lattice with bandwidth $W=4$. Time is measured in units of the inverse hopping, which is $W/4$ for the  Bethe lattice. The system  is initially prepared in the low-temperature anitferromagnetic phase at interaction $U_0=3$. The interaction is then suddenly increased to $U=8$, which destabilizes the antiferromagnetic order. Choosing the interaction quench as an excitation mechanism is rather arbitrary. We are mainly interested in the question how the time-resolved RIXS measurement reveals the dynamics of the order parameter, and therefore consider this setting which has already been studied in detail in the literature. In Fig.~\ref{fig06}b, the solid black line shows the  ultrafast decay of the N\'eel order parameter $m$ after the quench, where $m$ is defined as
\begin{align}
m
=
|n_{A,\uparrow}-n_{B,\uparrow}|=
|n_{A,\uparrow}-n_{A,\downarrow}|,
\end{align}
and $n_{j,\sigma}$ denotes the occupation of electrons with spin $\sigma$ on the sublattice $j=A,B$ of the bipartite lattice. Figure~\ref{fig06}a shows the RIXS signal for a probe pulse \eqref{rixsspulse} with duration $\tau=1$ and probe time $t_p=7$. For the chosen relatively short core-hole lifetime and probe duration, the spectrum does not depend strongly on the probe frequency $\omega_{\text{in}}$, so we tune $\omega_\text{in}=U/2=4$ to the broad intermediate-state resonance, and analyze the signal as a function of $\omega_\text{out}$. As for the impurity models in Figs.~\ref{fig04} and Fig.~\ref{fig05}, one observes a dominant elastic peak with $\omega_\text{out}=\omega_\text{in}$, and a loss peak around $\omega_\text{out}=\omega_\text{in}-U$. The comparison of the \CMethod{}  and \CFMethod{}  approaches shows that the role of vertex corrections to the signal is not too large in this regime, which may be expected for a system in the Mott phase. Quantitatively, the relevant loss peak is enhanced by about $20\%$ in the more accurate \CFMethod{}  approach. 

Figure \ref{fig06}b then analyzes the evolution of the loss peak $I_\text{loss}(t_p)=I_{\text{RIXS}}(\omega_{\text{in}},\omega_{\text{out}})$ at $\omega_{\text{in}}=4$, $\omega_{\text{out}}=\omega_{\text{in}}-U$ with probe time. The signal decreases, following  the decrease of the antiferromagnetic order parameter. If one normalizes the signal $I_{\text{loss}}(t_p)$ with the value at a given late time $t_p=12$, the prediction from the \CMethod{}  and the \CFMethod{}  approach basically fall on top of each other. This demonstrates how the decay of the antiferromagnetic order in the Mott phase can be extracted from the time-resolved RIXS signal.

\section{Conclusion}
\label{sec:conclusion} 

We have implemented a framework to evaluate the RIXS signal from the  quantum impurity model of non-equilibrium DMFT. The calculation is based on the solution of a time-dependent impurity model, and therefore avoids the need to compute a four-point correlation function in time, which would be hard to access even in exact diagonalizations of small clusters. We have formulated one approach that is based on the evaluation of a two-point response function in the driven model, and a direct evaluation of the RIXS  signal in terms of the outgoing photon occupation. The latter is numerically more costly, but formally includes vertex corrections that may be missing in the evaluation of the response function (depending on the method used to solve the impurity model).

This study opens a path to theoretically analyze the RIXS signal for time-resolved experiments on correlated systems. The DMFT framework self-consistently incorporates the time-dependent change of the local environment of a given site in the lattice which results from the non-equilibrium evolution. We have demonstrated this by probing the ultrafast dynamics of antiferromagnetic order in a Mott insulator. Future directions include in particular the application of the formalism to nonequilibrium studies of multi-orbital systems, which are hard to treat in direct cluster approaches. While the benchmark studies in the present paper are based on the NCA impurity solver, the general formalism can  be directly applied in combination with  different impurity solvers, such as higher-order variants of the hybridization expansion, or weak-coupling perturbation theory, in order to study systems outside the Mott regime. Another technically interesting question is whether the non-equilibrium DMFT bath can be represented with finitely many orbitals, as done in equilibrium\cite{Lu2019} or in Ref.~\onlinecite{Gramsch2013} for certain initial conditions. With this one could make use of the exact diagonalization formulation,\cite{Chen2019} while still capturing the self-consistent non-equilibrium evolution of the probe site environment provided by non-equilibrium DMFT.

\acknowledgements 
This work was supported by the ERC Starting Grant No.~716648 and ERC Consolidator Grant No.~724103. We acknowledge useful discussions with Philipp Hansmann and Steven Johnston.  
\begin{appendix}

\section{Kramers formula}

In this appendix, we recapitulate the derivation of the exact diagonalization formula \eqref{quasiexact} from the time-dependent result \eqref{gehejkl}. Using an expansion in eigenstates of the unperturbed electronic Hamiltonian, we first resolve the initial and final state in the electronic expectation value $\langle\cdots\rangle_{0}$. Performing an average over initial states $ |\Psi_i\rangle$ with weight $w_i$ and inserting an  identity $\sum_{f}      |\Psi_f\rangle\langle \Psi_f|  $ at the final time $t$, the expression for $\langle b^\dagger b \rangle_t$ factorizes as
\begin{align}
&\langle b^\dagger b \rangle_t
=
g^4\sum_{i,f} w_i |M_{i,f}|^2,
\\
&M_{i,f}=
 \sum_{\sigma}
\int_{t_0}^t dt_1 
\int_{t_0}^{t_1} dt_2 
\, s(t_2)
e^{-i\notilde \omega_{\text{in}}t_2+i\notilde \omega_{\text{out}} t_1}
\nonumber\\
&   \quad\quad\times \,\,e^{iE_f t_1} e^{-iE_it_2}   
 \langle \Psi_f| 
  P_{\text{out},\sigma}
   \mathcal{U}_0(t_1,t_2)
  P_{\text{in},\sigma}^\dagger
 |\Psi_i\rangle,
 \label{jjjkkkhhh}
\end{align}
where $E_i$ and $E_f$ are the initial and final state energies, respectively. We can now insert another identity $\sum_{m}  |\Psi_m\rangle\langle \Psi_m|  $ between the operators $P$ and $P^\dagger$,
\begin{align}
&M_{i,f}=
\sum_{m,\sigma}
\int_{t_0}^t \!\!dt_1 
\int_{t_0}^{t_1} \!\!dt_2 
\, s(t_2)
e^{-i\notilde \omega_{\text{in}}t_2+i\notilde \omega_{\text{out}} t_1}
e^{iE_f t_1-iE_it_2}  
 \nonumber
\\
&\times   \langle \Psi_f| 
  P_{\text{out},\sigma}
   |\Psi_m\rangle
e^{-i(E_m-i\Gamma)(t_1-t_2)}
   \langle \Psi_m|
  P_{\text{in},\sigma}^\dagger
 |\Psi_i\rangle.
 \label{gwgefgeje}
\end{align}
Here an ad-hoc approximation has been made, i.e., a lifetime $1/\Gamma$ has been added to the intermediate core state. Formally, the lifetime is due to the coupling to the (core) bath, which is taken into account explicitly in the  real-time simulations. We now evaluate the time integrals in Eq.~\eqref{gwgefgeje}, taking  $t$ and $t_0$ to $\pm\infty$, respectively. Rewriting the integral domain as $\int_{t_0}^\infty dt_2  \int_{t_2}^{\infty} dt_1$, and  substituting  $t_1-t_2 \to \bar t$ ($t_1\to t_2+\bar t$), the time integrals give
\begin{align}
&\int_{-\infty}^\infty \!\!dt_2 \,
 s(t_2)
 \int_{0}^{\infty} \!d\bar t\,
  e^{it_2(\notilde \omega_{\text{out}}-\notilde \omega_{\text{in}}+E_f-E_i)}
 \nonumber
 \\
&\,\,\,\times
  e^{i\bar t(\notilde \omega_{\text{out}} +E_f-E_m+i\Gamma)}
 =
\frac{i \notilde s(\notilde \omega_{\text{out}}-\notilde \omega_{\text{in}}+E_f-E_i)}
 {\notilde \omega_{\text{out}} +E_f-E_m+i\Gamma}, \label{gwgefgeje01}
\end{align}
where $\notilde s(\omega)$ is the Fourier transform of the pulse envelope. Inserting this expression into \eqref{gwgefgeje} gives Eq.~\eqref{quasiexact} of the main text.

\end{appendix}

%

\end{document}